\documentclass[iop,apj]{emulateapj}
\usepackage{graphicx,url}

\slugcomment{Accepted to ApJL on 2015 November 3}
\shorttitle{Triangulum II}
\shortauthors{Kirby et al.}
\begin{document}

\newcommand{\vmean}{-382.1}
\newcommand{\vmeanerr}{2.9}
\newcommand{\sigmav}{5.1}
\newcommand{\sigmaverr}{2.7}
\newcommand{\sigmaverrlower}{1.4}
\newcommand{\sigmaverrupper}{4.0}
\newcommand{\n}{6}
\newcommand{\fehmeanw}{-2.50}
\newcommand{\fehmeanwerr}{0.08}
\newcommand{\ml}{3600}
\newcommand{\mlerrlower}{2100}
\newcommand{\mlerrupper}{3500}
\newcommand{\smallsigmavjk}{2.8}
\newcommand{\smallsigmavjkerrl}{1.7}
\newcommand{\smallsigmavjkerru}{4.0}
\newcommand{\smallsigmavobjname}{65}
\newcommand{\sigmaverrjk}{1.2}
\newcommand{\varvdiff}{1.3}
\newcommand{\varvdifferr}{0.6}
\newcommand{\vgsr}{-262}
\newcommand{\density}{4.8}
\newcommand{\densityerrl}{3.5}
\newcommand{\densityerru}{ 8.1}

\title{Triangulum II: Possibly a Very Dense Ultra-Faint Dwarf Galaxy}

\author{Evan~N.~Kirby\altaffilmark{1},
  Judith~G.~Cohen\altaffilmark{1},
  Joshua~D.~Simon\altaffilmark{2},
  Puragra~Guhathakurta\altaffilmark{3}}

\altaffiltext{*}{The data presented herein were obtained at the
  W.~M.~Keck Observatory, which is operated as a scientific
  partnership among the California Institute of Technology, the
  University of California and the National Aeronautics and Space
  Administration. The Observatory was made possible by the generous
  financial support of the W.~M.~Keck Foundation.}
\altaffiltext{1}{California Institute of Technology, 1200 E.\ California Blvd., MC 249-17, Pasadena, CA 91125, USA}
\altaffiltext{2}{Observatories of the Carnegie Institution of Washington, 813 Santa Barbara Street, Pasadena, CA 91101, USA}
\altaffiltext{3}{UCO/Lick Observatory and Department of Astronomy and Astrophysics, University of California, 1156 High Street, Santa Cruz, CA 95064, USA}

\keywords{galaxies: dwarf --- Local Group --- galaxies: abundances}


\begin{abstract}

\citeauthor{lae15a}\ recently discovered Triangulum~II, a satellite of
the Milky Way.  Its Galactocentric distance is 36~kpc, and its
luminosity is only 450~$L_{\sun}$.  Using Keck/DEIMOS, we measured the
radial velocities of six member stars within 1.2\arcmin\ of the center
of Triangulum~II, and we found a velocity dispersion of $\sigma_v =
\sigmav_{-\sigmaverrlower}^{+\sigmaverrupper}$~km~s$^{-1}$.  We also
measured the metallicities of three stars and found a range of 0.8~dex
in [Fe/H]\@.  The velocity and metallicity dispersions identify
Triangulum~II as a dark matter-dominated galaxy.  The galaxy is moving
very quickly toward the Galactic center ($v_{\rm GSR} =
\vgsr$~km~s$^{-1}$).  Although it might be in the process of being
tidally disrupted as it approaches pericenter, there is no strong
evidence for disruption in our data set.  The ellipticity is low, and
the mean velocity, $\langle v_{\rm helio} \rangle = \vmean \pm
\vmeanerr$~km~s$^{-1}$, rules out an association with the
Triangulum--Andromeda substructure or the Pan-Andromeda Archaeological
Survey (PAndAS) stellar stream.  If Triangulum~II is in dynamical
equilibrium, then it would have a mass-to-light ratio of
$\ml_{-\mlerrlower}^{+\mlerrupper}~M_{\sun}~L_{\sun}^{-1}$, the
highest of any non-disrupting galaxy (those for which dynamical mass
estimates are reliable).  The density within the 3-D half-light radius
would be $\density_{-\densityerrl}^{+\densityerru}~M_{\sun}~{\rm
  pc}^{-3}$, even higher than Segue~1.  Hence, Triangulum~II is an
excellent candidate for the indirect detection of dark matter
annihilation.

\end{abstract}


\section{Introduction}
\label{sec:intro}

The Sloan Digital Sky Survey \citep[SDSS,][]{aba09} revolutionized
Local Group astronomy in the last decade by discovering more than a
dozen new dwarf galaxies around the Milky Way (MW)\@.  We are now in
the midst of another revolution.  The Panoramic Survey Telescope and
Rapid Response System \citep[Pan-STARRS,][]{kai10}, the Dark Energy
Survey \citep[DES,][]{fla12}, and other Dark Energy Camera (DECam)
imaging surveys have discovered more than 20 previously unknown MW
satellites \citep[e.g.,][]{lae15b,bec15,kim15e}.  The greater
photometric depth and expanded sky coverage of Pan-STARRS and DES over
SDSS has enabled the discovery of many new satellites with
luminosities less than $10^4~L_{\sun}$ and also satellites more
distant than 200~kpc.

Dwarf galaxy candidates are discovered through imaging, but their
identification as galaxies or star clusters is made secure through
spectroscopy \citep[e.g.,][]{wil12}.  A candidate can be considered a
galaxy if it shows evidence for dark matter, including a velocity
dispersion in excess of what would be expected from stellar mass alone
or a dispersion in stellar metallicity, which indicates chemical
self-enrichment.  Spectroscopy of satellites discovered in the last
two years has already confirmed five new galaxies and one globular
cluster \citep{sim15,wal15,kop15b,kir15b,mar15b}.

\citet{lae15a} discovered Triangulum~II (Tri~II) in Pan-STARRS images.
Its luminosity (450~$L_{\sun}$) and 2-D half-light radius (34~pc) are
comparable to Segue~1, the faintest galaxy known
\citep{bel07a,geh09,sim11}.  \citeauthor{lae15b}\ suggested that
Tri~II could be associated with the Triangulum--Andromeda halo
substructure \citep{maj04} or the Pan-Andromeda Archaeological Survey
(PAndAS) stream \citep{mar14}.  If so, then it could be one of the
progenitors of that tidal debris.  However, spatial coincidence is not
sufficient evidence for the association.  The velocity of the
progenitor should also match that of the debris.

We obtained spectra of stars in Tri~II in order to learn about its
origin and identity.  The velocity dispersion can identify it as a
galaxy or a star cluster, and the mean velocity can support or
disprove an association with stellar debris.  We describe our
observations in Section~\ref{sec:obs} and our measurements of
velocities and metallicities in Section~\ref{sec:spectroscopy}.  We
consider whether Tri~II is in dynamical equilibrium or tidally
disrupting in Section~\ref{sec:equilibrium}.  Finally, we discuss the
nature of Tri~II and its importance for the study of dark matter in
Section~\ref{sec:discussion}.


\section{Observations}
\label{sec:obs}

\subsection{Imaging}

\addtocounter{footnote}{-1}

We imaged Tri~II with Keck/LRIS \citep{oke95} on 2015 July 15.  We
obtained simultaneous 10~s exposures with $V$ and $I$ filters in the
blue and red channels, respectively.  We also obtained 10~s exposures
of the photometric standard field PG0231 in the same filters.  We
performed aperture photometry on both fields using SExtractor
\citep{ber96}.  The photometric zeropoint was determined by finding
the offsets between our instrumental magnitudes and P.B.~Stetson's
calibrated magnitudes in
PG0231.\footnote{\url{http://www.cadc-ccda.hia-iha.nrc-cnrc.gc.ca/en/community/
 STETSON/standards/}} We discarded resolved galaxies by eliminating
objects with $\texttt{class\_star} < 0.5$.

\subsection{Spectroscopy}

\begin{figure*}[t!]
\centering
\includegraphics[width=0.82\textwidth]{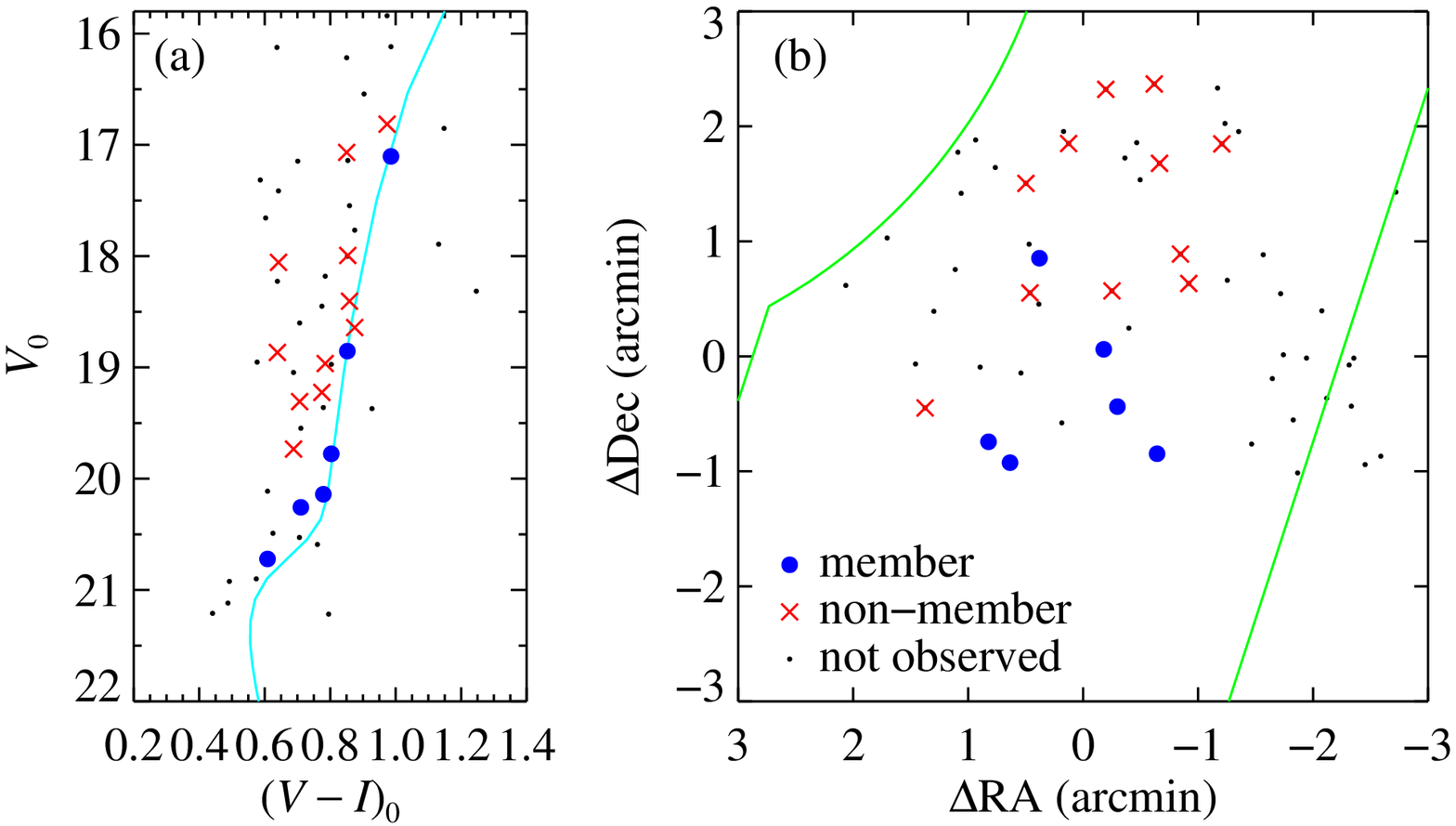}
\caption{({\textit a}) Color--magnitude diagram from LRIS photometry,
  showing spectroscopic members (large blue points) and non-members
  (red crosses).  The cyan line shows the ridgeline of the metal-poor
  globular cluster M92 \citep{cle06}.  ({\textit b}) The map of
  spectroscopic targets shown as distances from the center of
  Tri~II\@.  The DEIMOS slitmask outline is shown in green.  The full
  slitmask extends beyond the bounds of the figure because the field
  of view of LRIS---from which we selected DEIMOS targets---is smaller
  than for DEIMOS\@.\label{fig:cmd_map}}
\end{figure*}

We designed a slitmask for Keck/DEIMOS \citep{fab03} from the LRIS
photometry.  We selected 19 stars for spectroscopy based on their
locations in the color--magnitude diagram (CMD,
Figure~\ref{fig:cmd_map}a).  Stars near the sub-giant and RGB
tracks---the ``ridgeline''---of the metal-poor globular cluster M92
\citep{cle06} were considered for spectroscopy.  However, the field is
dense enough that we could not target every star.  When slitmask
design constraints forced a choice among several stars, we chose to
target the brightest star.  Figure~\ref{fig:cmd_map}b shows the 17
spectroscopic targets with S/N sufficient for velocity measurements
(see Section~\ref{sec:spectroscopy}).

We obtained 61~minutes of DEIMOS exposures in 1.6\arcsec\ seeing on
2015 October 6.  The poor seeing resulted in low S/N for the fainter
stars.  We obtained 52~minutes of exposures in 0.9\arcsec\ seeing on
2015 October 7.  The spectra from the first night were used only to
compare with velocity measurements from the second night
(Section~\ref{sec:rvmet}).  The measurements of velocity and
metallicity dispersions for Tri~II are based only on data from 2015
October 7.

We reduced the data with the \texttt{spec2d} software
\citep{coo12,new13} with modifications described by
\citet{kir15b,kir15a}.  Among other improvements, the 2-D wavelength
solution was improved by tracing the sky lines along the slit, and the
extraction was improved by taking into account differential
atmospheric refraction along the slit.


\section{Spectroscopic Measurements}
\label{sec:spectroscopy}

\subsection{Radial Velocities and Metallicities}
\label{sec:rvmet}

We measured heliocentric radial velocities ($v_{\rm helio}$) and
metallicities ([Fe/H]) in a manner identical to \citet{kir15b}, who
based their analysis on \citet{sim07}.  Radial velocities were
computed by finding the velocity that minimized the $\chi^2$ between
the observed spectrum and eight template spectra observed with
DEIMOS\@.  The template spectrum with the lowest $\chi^2$ was used.
We corrected the velocity shift due slit mis-centering by measuring
the observed wavelength of telluric absorption in the stellar spectrum
\citep[e.g.,][]{soh07}.  We computed errors due to random noise by
finding the standard deviation of the velocities of $10^3$ Monte Carlo
realizations of the spectrum.  The total error, $\delta v$, was
calculated by adding the random error in quadrature with a systematic
error of 1.49~km~s$^{-1}$.  The systematic error includes sources of
uncertainty that cannot be attributed to random noise, such as
uncorrected spectrograph flexure or small errors in the wavelength
solution.  The magnitude of the systematic error was calculated by
\citet{kir15b} by comparing repeated measurements of the same stars.

We tested our estimate of velocity errors by comparing the low-S/N
measurements of $v_{\rm helio}$ from 2015 October 6 to the high-S/N
measurements from 2015 October 7.  Of the six stars we determined to
be members (Section~\ref{sec:sigmav}), five velocities were measurable
with data from the first night.  We computed $(v_{\rm helio,1}-v_{\rm
  helio,2})/\sqrt{\delta v_1^2 + \delta v_2^2}$ for each pair of
measurements of the member stars.  The variance of this quantity
should be 1 if we estimated errors properly.  We measured it to be
$\varvdiff \pm \varvdifferr$.  Hence, the error estimates are
reasonable.

We measured effective temperatures ($T_{\rm eff}$) and metallicities
for member stars with sufficient S/N in the same manner as
\citet{kir08a,kir10}.  First, we divided the spectrum by a polynomial
fit to regions of the spectrum free of absorption lines.  Next, we
estimated temperatures and surface gravities ($\log g$) by fitting
Yonsei-Yale theoretical isochrones \citep{dem04} to observed stellar
colors and magnitudes.  Then, we used these parameters and an initial
guess of ${\rm [Fe/H]} = -1.5$ to construct a synthetic spectrum at
the observed spectrum's resolution.  This spectrum was linearly
interpolated from \citeauthor{kir10}'s (\citeyear{kir10}) synthetic
spectral grid.  In order to minimize $\chi^2$ between the observed and
synthetic spectra, we changed the synthetic spectrum's $T_{\rm eff}$
and [Fe/H] but held $\log g$ fixed.  The measured values of $T_{\rm
  eff}$ and [Fe/H] are those of the synthetic spectrum with the
minimum $\chi^2$.  The error on [Fe/H] is the appropriate diagonal
term of the covariance matrix added in quadrature with a systematic
error of 0.11~dex, which \citet{kir10} determined from repeat
measurements.  We kept [Fe/H] measurements of the three stars with
uncertainties less than 0.5~dex and discarded the others.

These three stars lie in the range $-3 < {\rm [Fe/H]} < -2$ with a
mean of $\langle {\rm [Fe/H]} \rangle = \fehmeanw \pm \fehmeanwerr$.
Tri~II has the lowest measured mean metallicity of any galaxy except
Segue~1 \citep{fre14} and Reticulum~II \citep{sim15,wal15}.  However,
the metallicity measurements for Tri~II are based on only three stars.
While the standard error of the mean is $\fehmeanwerr$~dex, the mean
metallicity of a larger sample could be substantially different.

\begin{deluxetable*}{lcccccr@{ }c@{ }lcccc}
\tablewidth{0pt}
\tablecolumns{11}
\tablecaption{Target List\label{tab:catalog}}
\tablehead{\colhead{ID} & \colhead{RA (J2000)} & \colhead{Dec (J2000)} & \colhead{$V_0$} & \colhead{$(V-I)_0$} & \colhead{S/N\tablenotemark{a}} & \multicolumn{3}{c}{$v_{\rm helio}$} & \colhead{Member?} & \colhead{$T_{\rm eff}$} & \colhead{$\log g$} & \colhead{[Fe/H]} \\
\colhead{ } & \colhead{ } & \colhead{ } & \colhead{(mag)} & \colhead{(mag)} & \colhead{(\AA$^{-1}$)} & \multicolumn{3}{c}{(km~s$^{-1}$)} & \colhead{ } & \colhead{(K)} & \colhead{(cm~s$^{-2}$)} & \colhead{ }}
\startdata
166                     & 02 13 11.42 & $+36$ 12 33.0 & 17.99 &            0.85 &          115 & $  11.0$ & $\pm$ & $ 5.9$ & N & \nodata & \nodata & \nodata \\
174                     & 02 13 12.85 & $+36$ 11 20.2 & 18.06 &            0.64 &          102 & $-100.8$ & $\pm$ & $ 1.6$ & N & \nodata & \nodata & \nodata \\
177                     & 02 13 13.21 & $+36$ 11 35.5 & 19.73 &            0.69 &     \phn  36 & $-280.7$ & $\pm$ & $ 2.5$ & N & \nodata & \nodata & \nodata \\
126                     & 02 13 14.11 & $+36$ 12 22.9 & 16.81 &            0.97 &          127 & $   2.4$ & $\pm$ & $ 1.5$ & N & \nodata & \nodata & \nodata \\
128                     & 02 13 14.21 & $+36$ 09 51.4 & 19.78 &            0.80 &     \phn  24 & $-384.9$ & $\pm$ & $ 3.2$ & Y & 5292 & 3.17 & $-2.72 \pm 0.39$ \\
127                     & 02 13 14.33 & $+36$ 13 04.3 & 19.31 &            0.71 &     \phn  54 & $ -86.0$ & $\pm$ & $ 1.8$ & N & \nodata & \nodata & \nodata \\
116                     & 02 13 15.92 & $+36$ 10 16.0 & 20.26 &            0.71 &     \phn  26 & $-377.6$ & $\pm$ & $ 3.7$ & Y & \nodata & \nodata & \nodata \\
113                     & 02 13 16.16 & $+36$ 11 16.3 & 18.97 &            0.78 &     \phn  72 & $ -57.6$ & $\pm$ & $ 1.6$ & N & \nodata & \nodata & \nodata \\
111                     & 02 13 16.42 & $+36$ 13 01.5 & 18.86 &            0.64 &     \phn  66 & $ -62.0$ & $\pm$ & $ 1.9$ & N & \nodata & \nodata & \nodata \\
106                     & 02 13 16.51 & $+36$ 10 45.9 & 17.10 &            0.99 &          219 & $-382.3$ & $\pm$ & $ 1.5$ & Y & 4922 & 1.88 & $-2.86 \pm 0.11$ \\
100                     & 02 13 18.03 & $+36$ 12 33.2 & 18.64 &            0.87 &     \phn  87 & $ -38.0$ & $\pm$ & $ 1.6$ & N & \nodata & \nodata & \nodata \\
91                      & 02 13 19.28 & $+36$ 11 33.4 & 20.14 &            0.78 &     \phn  29 & $-386.0$ & $\pm$ & $ 3.1$ & Y & \nodata & \nodata & \nodata \\
84                      & 02 13 19.69 & $+36$ 11 15.3 & 19.22 &            0.77 &     \phn  60 & $ -66.4$ & $\pm$ & $ 1.7$ & N & \nodata & \nodata & \nodata \\
82                      & 02 13 19.87 & $+36$ 12 12.4 & 18.41 &            0.86 &          101 & $-177.0$ & $\pm$ & $ 1.6$ & N & \nodata & \nodata & \nodata \\
76                      & 02 13 20.55 & $+36$ 09 46.7 & 20.72 &            0.61 &     \phn  17 & $-389.7$ & $\pm$ & $ 3.0$ & Y & \nodata & \nodata & \nodata \\
65                      & 02 13 21.48 & $+36$ 09 57.6 & 18.85 &            0.85 &     \phn  81 & $-374.5$ & $\pm$ & $ 1.7$ & Y & 5169 & 2.74 & $-2.04 \pm 0.13$ \\
45                      & 02 13 24.21 & $+36$ 10 15.3 & 17.07 &            0.85 &          152 & $ -62.1$ & $\pm$ & $ 1.5$ & N & \nodata & \nodata & \nodata \\
\enddata
\tablenotetext{a}{To convert to S/N per pixel, multiply by 0.57.}
\end{deluxetable*}

Table~\ref{tab:catalog} lists the radial velocities for all the stars
we observed with DEIMOS except two stars with spectra that were too
noisy to identify any absorption lines.  Temperatures and
metallicities are given for the three member stars with sufficient
quality for those measurements.

\subsection{Membership and Velocity Dispersion}
\label{sec:sigmav}

We identified a peak in the velocity distribution of the observed
stars around $-380$~km~s$^{-1}$.  We took stars within 50~km~s$^{-1}$
of this peak as the initial member list.  We measured the velocity
dispersion ($\sigma_v$) of these six stars in the same manner as
\citet{kir14,kir15b}, who based their analysis on \citet{wal06}.  We
estimated $\sigma_v$ via maximum likelihood.  A Monte Carlo Markov
chain (MCMC) with $10^7$ trials explored the parameter space of mean
velocity ($\langle v_{\rm helio} \rangle$) and $\sigma_v$.
We quote the values corresponding to the peaks of the probability
distributions as the measurements of $\langle v_{\rm helio} \rangle$
and $\sigma_v$.  The asymmetric $1\sigma$ confidence interval on
$\sigma_v$ is the range on either side of the mean value that bounds
68.3\% of the trials.

We measured $\langle v_{\rm helio} \rangle = \vmean \pm
\vmeanerr$~km~s$^{-1}$ and $\sigma_v =
\sigmav_{-\sigmaverrlower}^{+\sigmaverrupper}$~km~s$^{-1}$.  All six
candidate member stars are within $1.1\sigma_v$ of $\langle v_{\rm
  helio} \rangle$.  Furthermore, all six stars are close to the M92
ridgeline in Figure~\ref{fig:cmd_map}a, indicating that they pass a
CMD membership cut.  None of the six stars shows a strong Na~{\sc
  i}~8190 doublet, which would have indicated that the star is a
foreground dwarf.

\begin{figure}[t!]
\centering
\includegraphics[width=\columnwidth]{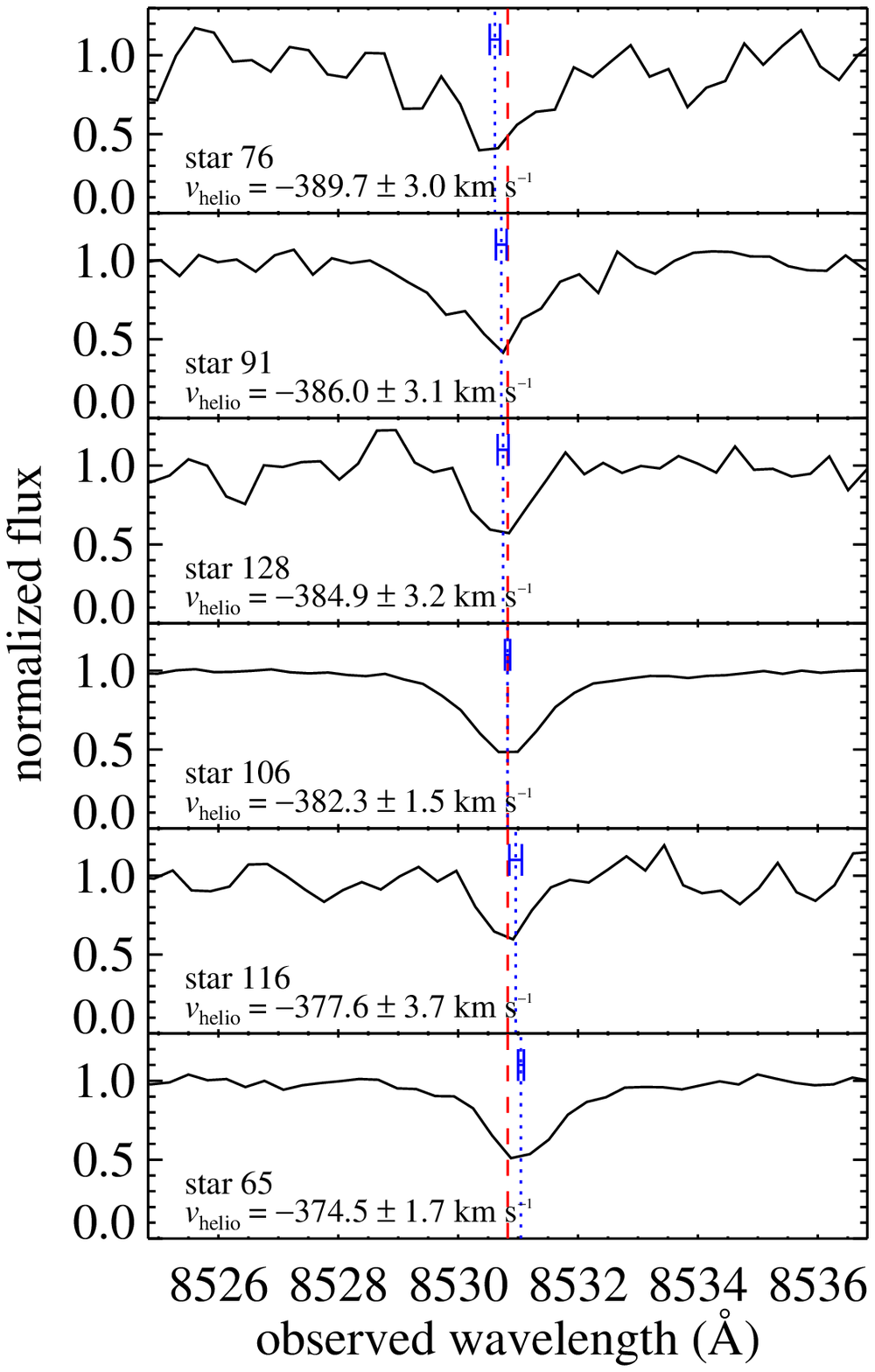}
\caption{Small regions of DEIMOS spectra of the six member stars shown
  against observed wavelength.  The full spectrum---much wider than
  shown here---is used for the velocity measurement.  The dashed red
  line shows the observed wavelength of Ca~{\sc ii}~8542 at the mean
  geocentric velocity of Tri~II\@.  The blue dotted lines show the
  observed wavelength of Ca~{\sc ii}~8542 for each star, and the blue
  whiskers indicate the $\pm 1\sigma$ uncertainty of the observed
  centroid of the absorption line for each star.  The spectra are
  ordered from the lowest to highest radial velocity.  The shift of
  Ca~{\sc ii}~8542 is apparent even by eye.\label{fig:spectra}}
\end{figure}

Figure~\ref{fig:spectra} shows the spectra of the six member stars in
observed wavelength.  Ca~{\sc ii}~8542 appears at a different observed
wavelength for every star, showing that we have resolved the velocity
dispersion of Tri~II.

Other than Tri~II, only three galaxies with $L < 10^4~L_{\sun}$ have
published measurements of $\sigma_v > 5$~km~s$^{-1}$: Bo{\" o}tes~II
\citep{koc09}, Pisces~II \citep{kir15b}, and Ursa Major~II
\citep{sim07}.  \citeauthor{koc09}\ measured $\sigma_v = 10.5 \pm
7.4$~km~s$^{-1}$ for Bo{\" o}tes~II, but more recent measurements have
found a smaller dispersion (M.~Geha et al., in prep.)\ and the
presence of at least one binary that inflates the apparent dispersion
\citep{ji15}.

Even if all of the stars are members, some of them might be binaries.
The orbital velocity of the binary would artificially inflate our
measurement of $\sigma_v$ for the galaxy.  We tested the robustness of
our measurement of $\sigma_v$ by jackknife resampling.  We
recalculated $\sigma_v$ for each of the six subsets of member stars
formed by removing one star.  All of the probability distributions are
well separated from zero.  The minimum velocity dispersion, calculated
by removing star \smallsigmavobjname, is
$\smallsigmavjk_{-\smallsigmavjkerrl}^{+\smallsigmavjkerru}$~km~s$^{-1}$.
The jackknife error, calculated as the standard deviation of
$\sigma_v$ for all of the jackknife trials, is
$\sigmaverrjk$~km~s$^{-1}$, somewhat smaller than the error calculated
from the MCMC distribution.


\section{Dynamical Equilibrium}
\label{sec:equilibrium}

\begin{deluxetable}{lr@{ }l}
\tablewidth{0pt}
\tablecolumns{3}
\tablecaption{Properties of Triangulum~II\label{tab:properties}}
\tablehead{\colhead{Property} & \multicolumn{2}{c}{Value}}
\startdata
$N_{\rm member}$ &  6 & \\
$\log (L_{V}/L_{\sun})$ & $2.65 \pm 0.20$ & \\
$r_h$ & $3.9_{-0.9}^{+1.1}$ & arcmin \\
$r_h$ & $ 34_{-  8}^{+  9}$ & pc \\
$\langle v_{\rm helio} \rangle$ & $-382.1 \pm 2.9$ & km~s$^{-1}$ \\
$v_{\rm GSR}$ & $-262$ km~s$^{-1}$ \\
$\sigma_v$ & $5.1_{-1.4}^{+4.0}$ & km~s$^{-1}$ \\
$\log (M_{1/2}/M_{\sun})$\tablenotemark{a} & $5.9_{-0.2}^{+0.4}$ & \\
$(M/L_V)_{1/2}$\tablenotemark{a,b} & $3600_{-2100}^{+3500}$ & $M_{\sun}~L_{\sun}^{-1}$ \\
$\rho_{1/2}$\tablenotemark{a,c} & $4.8_{-3.5}^{+ 8.1}$ & $M_{\sun}~{\rm pc}^{-3}$ \\
$\langle {\rm [Fe/H]} \rangle$ & $-2.50 \pm 0.08$ & \\
\enddata
\tablenotetext{a}{These quantities presume that Tri~II is in dynamical equilibrium.}
\tablenotetext{b}{Mass-to-light ratio within the half-light radius, calculated as $M_{1/2} = 4G^{-1}\sigma_v^2 r_h$ \citep{wol10}.}
\tablenotetext{c}{Density within the half-light radius.}
\tablerefs{The measurements of $\log L_V$ and $r_h$ come from \citet{lae15a}.}
\end{deluxetable}

Table~\ref{tab:properties} gives some characteristics of Tri~II,
including the mass within the 3-D half-light radius
\citep[$M_{1/2}$,][]{wol10}.  This quantity and its associated
quantities, mass-to-light ratio [$(M/L_V)_{1/2}$] and density
($\rho_{1/2}$) within the 3-D half-light radius, presume that the
galaxy is spherically symmetric and in dynamical equilibrium.
However, the velocity dispersion accurately reflects the mass even in
the presence of moderate tidal forces \citep{oh95}.  If the velocities
of the stars we measured are very heavily affected by tides, then
these quantities are not meaningful.  We now consider whether the
center of Tri~II is in dynamical equilibrium.

The MW exerts the maximum tidal shear on satellite galaxies at their
pericenters \citep[e.g.,][]{may01}.  It is more likely to find a
tidally disrupting galaxy close to the Galactic center than far from
it.  Tri~II is only $D_{\rm GC} = 36 \pm 2$~kpc from the Galactic
center \citep{lae15a}.  It is also rapidly approaching its pericenter.
Assuming a solar orbital velocity of 220~km~s$^{-1}$, the velocity of
Tri~II relative to the Galactic standard of rest (GSR) is
$\vgsr$~km~s$^{-1}$.  The fact that Tri~II is approaching pericenter
rather than receding from it is consistent with its imminent tidal
disruption.  However, its large velocity limits the time that it will
spend near pericenter and consequently reduces the total tidal effect.

A tidally disrupting galaxy could have a high ellipticity.  The
ellipticity of Tri~II is $\epsilon = 0.21_{-0.21}^{+0.17}$
\citep{lae15a}.  In contrast with presently disrupting galaxies, like
Sagittarius \citep[$\epsilon = 0.65$,][]{maj03}, Tri~II is not
obviously elliptical.

Along with high ellipticity, ongoing tidal disruption could cause a
non-Gaussian velocity distribution.  A Shapiro-Wilk test gives a $p$
value of 0.87.  A completely Gaussian distribution would have a $p$
value of 1.  Therefore, there is no evidence for non-Gaussianity in
the velocity distribution.  Of course, our small sample size limits
the significance of this result.  Furthermore, we have measured only
line-of-sight velocities, and even a tidally disrupting system may
have normally distributed velocities along some lines of sight.

We estimated a lower limit to the tidal radius assuming that $M_{1/2}$
is the entire mass of Tri~II\@.  The Roche limit for a fluid satellite
is $r_{\rm tidal} \sim 0.4\,D_{\rm GC}(M_{1/2}/M_{\rm MW})^{1/3}$,
where $M_{\rm MW} \approx 10^{12}~M_{\sun}$ is the MW's mass.  Under
these assumptions, $r_{\rm tidal} \approx 140$~pc for Tri~II, or about
three times its 3-D half-light radius ($4/3$ of the 2-D, projected
half-light radius).  The tidal radius would shrink to the same value
as the 3-D half-light radius at $\sim 12$~kpc from the Galactic
center.  Therefore, all of the stars we observed in Tri~II are
presently insulated from Galactic tides.  Although this estimate of
tidal radius presumes that the velocity dispersion reflects the
present mass, simulations suggest that the velocity dispersion is a
good indicator of the instantaneous mass except for a short time after
pericenter, even for systems experiencing significant tidal stripping
\citep{oh95,mun08,pen09}.

\citet{lae15a} noted the possible association of Tri~II with the
Triangulum--Andromeda halo substructure \citep{maj04} or the PAndAS
stream \citep{mar14}.  Association with such halo debris might
indicate that the galaxy is being disrupted and that it is the source
of the debris.  However, \citet{dea14} measured the GSR velocities of
these structures as 30--70~km~s$^{-1}$, which is roughly
300~km~s$^{-1}$ different from $v_{\rm GSR}$ for Tri~II\@.  Therefore,
there is no presently known stream that could be associated with
Tri~II\@.  This does not prove that Tri~II is in dynamical
equilibrium, but it does show that, if Tri~II is being tidally
disrupted, it is not the source of the Triangulum--Andromeda or PAndAS
stellar debris.

\begin{figure}[t!]
\centering
\includegraphics[width=\columnwidth]{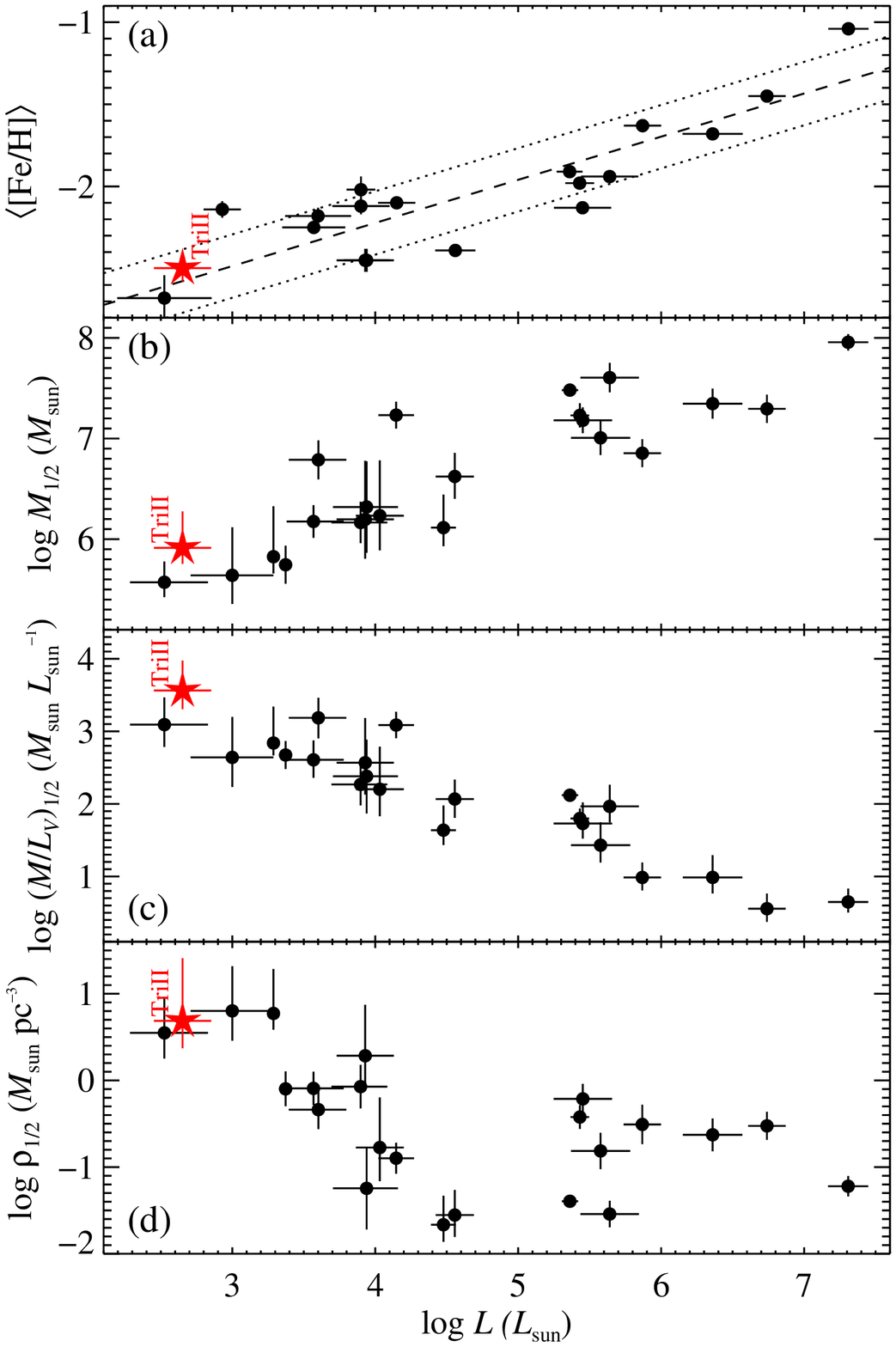}
\caption{({\textit a}) Luminosity--metallicity relation for MW
  satellite galaxies.  The dashed line shows the linear fit to the
  galaxies except Tri~II \citep{kir13b,kir15b,fre14}, and the dotted
  lines show the rms dispersion about the fit.  ({\textit b}) Masses
  of MW satellite galaxies within their 3-D half-light radii assuming
  dynamical equilibrium.  Data are from \citet[][and references
    therein]{mcc12}, \citet{sim15}, \citet{kop15b}, and
  \citet{kir15b}.  ({\textit c}) Mass-to-light ratios and ({\textit
    d}) densities within the 3-D half-light radii.  Bo{\" o}tes~II and
  III are not shown because their published velocity dispersions do
  not reflect their dynamical masses.\label{fig:trends}}
\end{figure}

The luminosity--metallicity relation (LZR, Figure~\ref{fig:trends}a)
is a diagnostic of tidal stripping.  The LZR for classical dwarf
galaxies is very tight, with an rms of only 0.13~dex
\citep{kir11a,kir13b}.  If a galaxy initially conforms to the LZR, then
tidal stripping will decrease its luminosity while keeping its average
metallicity roughly constant.  This corresponds to a leftward move in
Figure~\ref{fig:trends}a.  Some of the galaxies with $L \lesssim
10^4~L_{\sun}$, especially Segue~2 \citep{kir13a}, lie significantly
to the left of the LZR\@.  On the other hand, Tri~II is consistent
with the LZR\@.  Therefore, any tidal stripping that already happened
is likely to have been mild.


\section{Discussion}
\label{sec:discussion}

Tri~II satisfies the definition of ``galaxy'' given by \citet{wil12}.
The velocity dispersion is much too large to be explained by stars
alone.  We also found a large dispersion in metallicity.  The stars
span 0.8~dex in [Fe/H], which is evidence for chemical
self-enrichment.  The present mass of stars alone would not have been
enough to retain supernova ejecta.  Hence, without substantial mass
loss, the velocity and metallicity dispersions are evidence for a
large amount of dark matter.

It is unclear whether Tri~II is in dynamical equilibrium.  With a
total luminosity of only 450~$L_{\sun}$, the galaxy has very few stars
available to measure its shape very precisely.  Even fewer stars are
available for spectroscopy.  Therefore, resolving this question will
be very difficult.  Regardless, we now consider Tri~II's place among
the MW satellite population under the presumption of dynamical
equilibrium.

Figures~\ref{fig:trends}b--d show the trends of $M_{1/2}$,
$(M/L_V)_{1/2}$, and $\rho_{1/2}$ with luminosity.  Tri~II has the
largest mass-to-light ratio
($\ml_{-\mlerrlower}^{+\mlerrupper}~M_{\sun}~L_{\sun}^{-1}$) of any
galaxy except Bo{\" o}tes~III, whose tidal disruption is nearly
complete \citep{gri09,car09}.  If Tri~II is in dynamical equilibrium,
then it is the most dark-matter dominated galaxy known.

The five galaxies with $\rho_{1/2} > 1~M_{\sun}~{\rm pc}^{-3}$ in
order from densest to least dense are Willman~1, Horologium~I, Tri~II,
Segue~1, and Pisces~II\@.  Four of these galaxies comprise the least
luminous galaxies with measured velocity dispersions.  This
correlation could arise because these galaxies also have the smallest
half-light radii.  If galaxies' mass profiles peak in the center, then
smaller galaxies will be observed to have larger $\rho_{1/2}$, even if
their total masses and mass profiles are identical.  The correlation
between $\rho_{1/2}$ and $L_V$ could also arise because less massive
galaxies are more susceptible to tidal stripping.  Hence, the
measurement of $\rho_{1/2}$ would be invalid because the galaxies are
not in equilibrium.  Willman~1 has a velocity distribution that does
not seem consistent with dynamical equilibrium \citep{wil11}.  The
velocity dispersions of Horologium~I, Pisces~II, and Tri~II were all
measured from 5--7 stars \citep[][and this work]{kop15b,kir15b}.
Hence, Segue~1 \citep{sim11} remains the galaxy with the most secure
measurement of a very high central density.

In summary, we measured $\sigma_v =
\sigmav_{-\sigmaverrlower}^{+\sigmaverrupper}$~km~s$^{-1}$ for
Tri~II\@.  The present measurements cannot determine whether the
galaxy is in dynamical equilibrium or being tidally disrupted.
However, the possibility that it is in equilibrium is very exciting.
Tri~II would be the most dark-matter dominated galaxy known, and it
would be an excellent candidate for the indirect, gamma-ray detection
of dark matter annihilation.  The annihilation signal scales as
$\rho^2$, which makes very dense galaxies---possibly including
Tri~II---the best prospects for detection.

\acknowledgments We thank Gina Duggan for obtaining LRIS images, Emily
Cunningham for helpful statistics advice, and the anonymous referee
for helpful feedback.  PG acknowledges support from NSF grants
AST-1010039 and AST-1412648.  We are grateful to the many people who
have worked to make the Keck Telescope and its instruments a reality
and to operate and maintain the Keck Observatory.  The authors wish to
extend special thanks to those of Hawaiian ancestry on whose sacred
mountain we are privileged to be guests.  Without their generous
hospitality, none of the observations presented herein would have been
possible.

{\it Facility:} \facility{Keck:I (LRIS), Keck:II (DEIMOS)}

\end{document}